\documentclass[preprint,onecolumn,aps,pre,eqsecnum,showpacs,showkeys,a4paper]{revtex4-1}
\usepackage[latin9]{inputenc}
\usepackage{color}
\definecolor{note_fontcolor}{rgb}{0.80078125, 0.80078125, 0.80078125}
\usepackage[english]{babel}
\usepackage{amsmath}
\usepackage{amssymb}
\usepackage{graphicx}
\usepackage{esint}
\usepackage[unicode=true,
 bookmarks=true,bookmarksnumbered=true,bookmarksopen=true,bookmarksopenlevel=1,
 breaklinks=true,pdfborder={0 0 0},backref=false,colorlinks=true]
 {hyperref}
\hypersetup{pdftitle={Modeling double slit interference via anomalous diffusion},
 pdfauthor={Johannes Mesa Pascasio}}

\makeatletter

\newenvironment{lyxgreyedout}
  {\textcolor{note_fontcolor}\bgroup\ignorespaces}
  {\ignorespacesafterend\egroup}

\@ifundefined{textcolor}{}
{%
 \definecolor{BLACK}{gray}{0}
 \definecolor{WHITE}{gray}{1}
 \definecolor{RED}{rgb}{1,0,0}
 \definecolor{GREEN}{rgb}{0,1,0}
 \definecolor{BLUE}{rgb}{0,0,1}
 \definecolor{CYAN}{cmyk}{1,0,0,0}
 \definecolor{MAGENTA}{cmyk}{0,1,0,0}
 \definecolor{YELLOW}{cmyk}{0,0,1,0}
 }
\numberwithin{equation}{section}
\numberwithin{figure}{section}
\numberwithin{table}{section}

\usepackage{graphicx}

\DeclareGraphicsExtensions{.png,.pdf,.eps}
\graphicspath{{pictures/},{.}}

\@ifundefined{showcaptionsetup}{}{%
 \PassOptionsToPackage{caption=false}{subfig}}
\usepackage{subfig}
\makeatother

\begin{document}

\title{Modeling double slit interference via anomalous diffusion: independently
variable slit widths}

\author{Johannes \surname{Mesa Pascasio}\textsuperscript{1,2}}

\email[E-mail: ]{ains@chello.at}

\homepage[Visit: ]{http://www.nonlinearstudies.at/}

\author{Siegfried \surname{Fussy}\textsuperscript{1}}

\email[E-mail: ]{ains@chello.at}

\homepage[Visit: ]{http://www.nonlinearstudies.at/}

\author{Herbert \surname{Schwabl}\textsuperscript{1}}

\email[E-mail: ]{ains@chello.at}

\homepage[Visit: ]{http://www.nonlinearstudies.at/}

\author{Gerhard \surname{Grössing}\textsuperscript{1}}

\email[E-mail: ]{ains@chello.at}

\homepage[Visit: ]{http://www.nonlinearstudies.at/}

\affiliation{\textsuperscript{1}Austrian Institute for Nonlinear Studies, Akademiehof\\
 Friedrichstr.~10, 1010 Vienna, Austria}

\affiliation{\textsuperscript{2}Institute for Atomic and Subatomic Physics, Vienna
University of Technology\\
Operng.~9, 1040 Vienna, Austria \vspace*{4cm}
}

%\date{\today}
\begin{abstract}
Based on a re-formulation of the classical explanation of quantum
mechanical Gaussian dispersion (Grössing~\textit{et~al}. 2010 \cite{Groessing.2010emergence})
as well as interference of two Gaussians (Grössing~\textit{et~al}.
2012 \cite{Groessing.2012doubleslit}), we present a new and more
practical way of their simulation. The quantum mechanical ``decay
of the wave packet'' can be described by anomalous sub-quantum diffusion
with a specific diffusivity varying in time due to a particle's changing
thermal environment. In a simulation of the double-slit experiment
with different slit widths, the phase with this new approach can be
implemented as a local quantity. We describe the conditions of the
diffusivity and, by connecting to wave mechanics, we compute the exact
quantum mechanical intensity distributions, as well as the corresponding
trajectory distributions according to the velocity field of two Gaussian
wave packets emerging from a double-slit. We also calculate probability
density current distributions, including situations where phase shifters
affect a single slit's current, and provide computer simulations thereof.%

\begin{lyxgreyedout}
\global\long\def\VEC#1{\mathbf{#1}}
\global\long\def\d{\,\mathrm{d}}
\global\long\def\e{{\rm e}}
\global\long\def\meant#1{\left<#1\right>}
\global\long\def\meanx#1{\overline{#1}}
\global\long\def\mpbracket{\ensuremath{\genfrac{}{}{0pt}{1}{-}{\scriptstyle (\kern-1pt +\kern-1pt )}}}
\global\long\def\pmbracket{\ensuremath{\genfrac{}{}{0pt}{1}{+}{\scriptstyle (\kern-1pt -\kern-1pt )}}}
\global\long\def\p{\partial}
\end{lyxgreyedout}

\end{abstract}


\maketitle

\section{Introduction}

In reference \cite{Groessing.2010emergence} we presented a classical
model for the explanation of quantum mechanical dispersion of a free
Gaussian wave packet. In accordance with the classical model, we shall
now relate it more directly to a ``double solution'' analogy gleaned
from \cite{Couder.2012probabilities}. For, as is shown, e.g., in
\cite{Holland.1993,Elze.2011general}, one can construct various forms
of classical analogies to quantum mechanical Gaussian dispersion.
Originally, the expression of a ``double solution'' refers to an
early idea of \cite{DeBroglie.1960book} to model quantum behavior
by a two-fold process, i.e., by a the movement of a hypothetical point-like
``singularity solution'' of the Schrödinger equation, and by the
evolution of the usual wave function that would provide the empirically
confirmed statistical predictions. Recently, \cite{Couder.2012probabilities}
used this ansatz to describe the behaviors of their ``bouncer''-
(or ``walker''-) droplets: On an individual level, one observes
particles surrounded by circular waves they emit through the phase-coupling
with an oscillating bath, which provides, on a statistical level,
the emergent outcome in close analogy to quantum mechanical behavior
(like, e.g., diffraction or double-slit interference). The simulation
of interference in the double-slit experiment was in \cite{Groessing.2012doubleslit}
easily achieved by assuming the simple case where the two slits have
equal aperture. Instead, in this paper, we discard that simplification
and show in a more detailed analysis that one can i) find a formulation
applicable to independently variable slit widths, and ii) provide
computer simulations thereof.

\section{From classical phase-space distributions to quantum mechanical dispersion}

In the context of the double solution idea, which is related to correlations
on a statistical level between individual uncorrelated particle positions
$x$ and momenta $p$, respectively, we consider the free Liouville
equation
\begin{equation}
\frac{\p f}{\p t}+\sum_{i=1}^{3}\frac{p_{i}}{m}\frac{\p f}{\p x_{i}}-\sum_{i=1}^{3}\frac{\p V}{\p x_{i}}\frac{\p f}{\p p_{i}}=0\label{eq:td.1}
\end{equation}
with potential $V$ and mass $m$. For simplicity, we restrict ourselves
to the 1-dimensional space coordinate $x$ further on. Liouville's
equation~(\ref{eq:td.1}) implies the spatial conservation law and
has the property that precise knowledge of initial conditions is not
lost in the course of time. That is, it provides a phase-space distribution
$f\left(x,p,t\right)$ that shows the emergence of correlations between
$x$ and $p$ from an initially uncorrelated product function of non-spreading
(``classical'') Gaussian position distributions as well as momentum
distributions,
\begin{equation}
f_{0}\left(x,p\right)=\frac{1}{2\pi\sigma_{0}\pi_{0}}\exp\left\{ -\frac{x^{2}}{2\sigma_{0}^{2}}\right\} \exp\left\{ -\frac{p^{2}}{2\pi_{0}^{2}}\right\} ,\label{eq:td.2}
\end{equation}
where $\sigma_{0}$ is the initial space deviation, i.e., $\sigma_{0}=\sigma(t=0)$,
and $\pi_{0}:=mu_{0}$ is the momentum deviation. Then the phase-space
distribution reads as
\begin{equation}
f\left(x,p,t\right)=\frac{1}{2\pi\sigma_{0}mu_{0}}\exp\left\{ -\frac{\left(x-pt/m\right)^{2}}{2\sigma_{0}^{2}}\right\} \exp\left\{ -\frac{p^{2}}{2m^{2}u_{0}^{2}}\right\} .\label{eq:td.3}
\end{equation}

\noindent The above-mentioned correlations between $x$ and $p$ emerge
when one considers the probability density in $x$--space, which is
given by the integral
\begin{equation}
P\left(x,t\right)=\int f\d p=\frac{1}{\sqrt{2\pi}\sigma}\exp\left\{ -\frac{x^{2}}{2\sigma^{2}}\right\} ,\label{eq:td.4}
\end{equation}
with the variance at time $t$ given by
\begin{equation}
\sigma^{2}=\sigma_{0}^{2}+u_{0}^{2}\, t^{2}.\label{eq:td.5}
\end{equation}

In other words, the distribution~(\ref{eq:td.4}) describing a spreading
Gaussian is obtained from a continuous set of classical, non-spreading,
Gaussian position distributions whose momenta also have a non-spreading
Gaussian distributions. One thus obtains the exact quantum mechanical
dispersion formula for a Gaussian, as we have obtained also previously
from our classical ansatz by relating different kinetic energy terms
in our diffusion model~\cite{Groessing.2010emergence}. For confirmation
with respect to that model we use the Einstein relation
\begin{equation}
D=\frac{\hbar}{2m}\,,\label{eq:td.6}
\end{equation}
with the reduced Planck constant $\hbar=h/(2\pi)$ and $m$ being
the particle's mass, and we note that with (\ref{eq:td.4}), $\nabla P=\frac{\p}{\p x}P=-\frac{x}{\sigma^{2}}P$
and the usual definition of the ``osmotic'' velocity $u$ one obtains
\begin{equation}
u=u(x,t)=-D\frac{\nabla P}{P}=\frac{xD}{\sigma^{2}}\,.\label{eq:td.7}
\end{equation}
For the average initial value we find
\begin{equation}
u_{0}:=\left.\vphantom{\int}\meanx u\right|_{t=0}=u(\sigma_{0},0)=\frac{D}{\sigma_{0}}\,,\label{eq:td.8}
\end{equation}
which turns out to be the same as the initial velocity at position
$x=\sigma_{0}$ at $t=0$. This is a characteristic value for Gaussians,
which we simply called ``initial velocity'' in our recent papers.
It corresponds exactly to the velocity $u$ at starting time $t=0$
at the trajectory that has distance $\xi(0)=\pm\sigma_{0}$ from the
maximum of the Gaussian (Fig.~\ref{fig:td.1}). With Eq.~(\ref{eq:td.8})
one can rewrite Eq.~(\ref{eq:td.5}) in the more familiar form
\begin{equation}
\sigma^{2}=\sigma_{0}^{2}\left(1+\frac{D^{2}t^{2}}{\sigma_{0}^{4}}\right).\label{eq:td.9}
\end{equation}

\noindent Note also that by using the Einstein relation~(\ref{eq:td.6})
the norm in~(\ref{eq:td.3}) becomes the invariant expression

\noindent 
\begin{equation}
\frac{1}{2\pi\sigma_{0}mu_{0}}=\frac{1}{2\pi mD}=\frac{2}{h}\label{eq:td.10}
\end{equation}
reflecting the ``exact uncertainty relation''~\cite{Hall.2002schrodinger}.

\section{Spreading of the wave packet due to a path excitation field}

\noindent We note that $\sigma/\sigma_{0}$ is a spreading ratio for
the wave packet independent of $x$. This functional relationship
is thus not only valid for the particular point $\xi(t)=\sigma(t)$,
but for all $x$ of the Gaussian. Therefore, one can generalize (\ref{eq:td.9})
for all $x$, i.e., 
\begin{equation}
\xi(t)=\xi(0)\frac{\sigma}{\sigma_{0}},\qquad\text{where }\quad\frac{\sigma}{\sigma_{0}}=\sqrt{1+\frac{D^{2}t^{2}}{\sigma_{0}^{4}}}\;.\label{eq:td.11}
\end{equation}
In other words, one derives also the time-invariant ratio for the
spreading 
\begin{equation}
\frac{\xi(t)}{\sigma}=\frac{\xi(0)}{\sigma_{0}}=\text{const.}\label{eq:td.12}
\end{equation}
\begin{figure}[th]
\centering{}\includegraphics{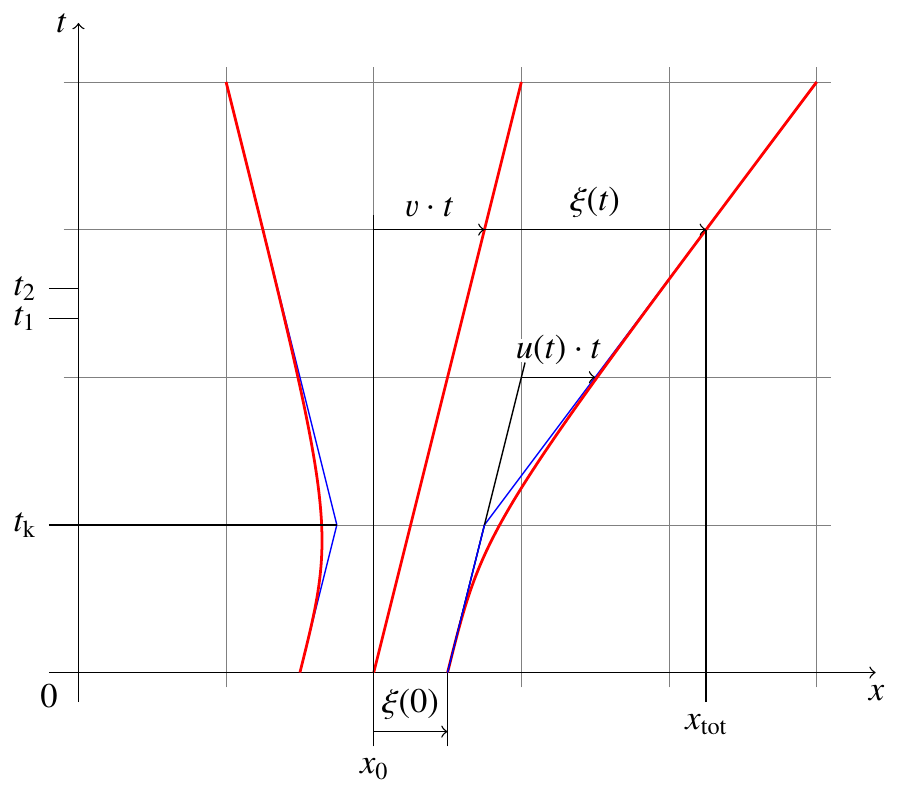}\caption{Bohm-type trajectories for a quantum particle with initial Gaussian
distribution exhibiting the characteristics of ballistic diffusion\label{fig:td.1}}
\end{figure}
In Fig.~\ref{fig:td.1} the spreading according to Eq.~(\ref{eq:td.11})
is sketched. We can now try to implement our previous assumption that
the ``bouncer'' particle is phase locked with its nonlocal diffusion
wave field such that the Gaussian describing the diffusion process
has long undulatory tails representing the locking in with the undulations
of the zero-point field. In other words, we can now re-consider our
classical simulations of Gaussian dispersion and double slit interference
\cite{Groessing.2012doubleslit}, respectively, by constructing from
(\ref{eq:td.4}) a description of our ``path excitation field''
via the introduction of the amplitude $R$ as product of a Gaussian
(at rest in the $x$--direction) and a plane wave (in the $y$--direction),
\begin{equation}
R\left(x,t\right)=\left(2\pi\sigma^{2}\right)^{-1/4}\exp\left\{ -\frac{x^{2}}{4\sigma^{2}}\right\} \cos\left(k_{y}y\right).\label{eq:td.13}
\end{equation}
The product (\ref{eq:td.13}) is factorisable for all $t$ into $x$--
and $y$--dependent functions, due to the motion in the $y$--direction
by $y\left(t\right)=\hbar kt/m$. According to our principle of path
excitation \cite{Groessing.2012doubleslit}, we deal with a single,
classical particle of velocity $v=p/m$ following the propagations
of waves of equal amplitude $R$ comprising a wave-like thermal bath
that permanently provides some momentum fluctuations $\delta p$. 

The latter are reflected in Eq.~(\ref{eq:td.9}) via the r.m.s.~deviation
$\sigma(t)$ from the classical path. In other words, one has to do
with a wave packet with an overall uniform velocity $v$, where the
position $x_{0}=vt$ moves like a free classical particle, as indicated
in Fig.~\ref{fig:td.1}. As the packet spreads according to Eq.~(\ref{eq:td.9}),
$\xi(t)$ describes the result of the motion along a trajectory of
a point of this packet that was initially at $\xi(0)$. The smaller
the initial value of $\left|\xi(0)\right|$, i.e., the distance from
$x_{0}$ of the center point of the packet, the slower said spreading
takes place. In our model picture, this is easy to understand: For
a particle exactly at the center of the packet, $x_{{\rm tot}}=x_{0}\Leftrightarrow\xi(0)=0$
, the momentum contributions from the ``heated up'' environment
on average cancel each other for symmetry reasons. However, the further
off a particle is from that center, the stronger this symmetry will
be broken, i.e., leading to a position-dependent net acceleration
or deceleration, respectively, or, in effect, to the ``decay of the
wave packet''. The actual decay of the wave packet starts, roughly
spoken, at a time $t_{{\rm k}}$, indicated by a ``kink'' in Fig.~\ref{fig:td.1}
which is due to the squared time-behavior in Eq.~(\ref{eq:td.9}). 

From Fig.~\ref{fig:td.1} we find $x_{{\rm tot}}=x_{0}+v(t)t+\xi(t)$
and $\xi(t)=\xi(0)+u(t)t.$ Without loss of generality we set $v={\rm const.}$
and $x_{0}=0$ further on. With the use of Eq.~(\ref{eq:td.11})
we obtain 
\begin{equation}
x_{{\rm tot}}(t)=vt+\xi(t)=vt+\xi(0)\frac{\sigma}{\sigma_{0}}=vt+\xi(0)\sqrt{1+\frac{u_{0}^{2}t^{2}}{\sigma_{0}^{2}}}\;.\label{eq:td.14}
\end{equation}
In our model picture, $x_{{\rm tot}}$ is the position of the ``smoothed
out'' \textit{trajectories}, i.e., those averaged over a very large
number of Brownian motions. 

Moreover, one can now also calculate the \textit{average total velocity
field of a Gaussian wave packet} as 
\begin{equation}
v_{{\rm tot}}(t)=\frac{\d x_{{\rm tot}}(t)}{\d t}=v+\xi(0)\,\frac{u_{0}^{2}t/\sigma_{0}^{2}}{\sqrt{1+u_{0}^{2}t^{2}/\sigma_{0}^{2}}}\;,\label{eq:td.15}
\end{equation}
which describes the velocity field $v_{{\rm tot}}$ of a point along
a trajectory (i.e, the residue of the ``path excitation field''
to be explicated further below).

Finally, we derive the \textit{average total acceleration field of
a Gaussian wave packet} as 
\begin{equation}
a_{{\rm tot}}(t)=\frac{\d v_{{\rm tot}}(t)}{\d t}=\xi(0)\,\frac{u_{0}^{2}/\sigma_{0}^{2}}{\sqrt{\left(1+u_{0}^{2}t^{2}/\sigma_{0}^{2}\right)^{3}}}\;,\label{eq:td.16}
\end{equation}
describing the acceleration of a point along the trajectory at time
$t$. Eqs.~(\ref{eq:td.14}) to (\ref{eq:td.16}) allow us to calculate
the quantities along a trajectory only from a given starting point,
indicated by $\xi(0)$.

Actually, however, we are interested in the dynamics at any given
position $(x,t)$ directly. Using 
\begin{equation}
\xi(t)=x-vt\label{eq:td.17}
\end{equation}
 and Eq.~(\ref{eq:td.11}) we rewrite
\begin{equation}
\xi(0)=\frac{x-vt}{\sqrt{1+u_{0}^{2}t^{2}/\sigma_{0}^{2}}}
\end{equation}
which leads to the generalized fields,
\begin{align}
x_{{\rm tot}}(x,t) & =x,\vphantom{\intop_{0}^{0}}\label{eq:td.19}\\
v_{{\rm tot}}(x,t) & =v+\xi(t)\,\frac{u_{0}^{2}t/\sigma_{0}^{2}}{1+u_{0}^{2}t^{2}/\sigma_{0}^{2}}=v+(x-vt)\,\frac{u_{0}^{2}t}{\sigma^{2}}\,,\vphantom{\intop_{0}^{0}}\label{eq:td.20}\\
a_{{\rm tot}}(x,t) & =\xi(t)\,\frac{u_{0}^{2}/\sigma_{0}^{2}}{\left(1+u_{0}^{2}t^{2}/\sigma_{0}^{2}\right)^{2}}=(x-vt)\,\frac{u_{0}^{2}\sigma_{0}^{2}}{\sigma^{4}}\;,\vphantom{\intop_{0}^{0}}\label{eq:td.21}
\end{align}
which will be used in the simulations later on.

\section{The derivation of $D_{{\rm t}}$}

We derive a solution for a diffusion equation with a time-dependent
diffusion coefficient $kt^{\alpha}$ for a generalized diffusion equation
(cf. \cite{Mesa.2012classical}) 
\begin{equation}
\frac{\partial P}{\partial t}=kt^{\alpha}\frac{\partial^{2}P}{\partial x^{2}}\;,\quad\alpha>0.\label{eq:tdde.1}
\end{equation}
Here, $t$ and $k$ denote the time and a constant factor, respectively.
Inserting $P(x,t)$ of Eq.~(\ref{eq:td.4}) as a solution into Eq.~(\ref{eq:tdde.1})
yields 
\begin{align}
\frac{P\dot{\sigma}}{\sigma}\left(\frac{x^{2}}{\sigma^{2}}-1\right) & =kt^{\alpha}\frac{P}{\sigma^{2}}\left(\frac{x^{2}}{\sigma^{2}}-1\right)\;,
\end{align}
and, after integration, 
\begin{align}
\frac{\sigma^{2}}{2} & =k\frac{t^{\alpha+1}}{\alpha+1}+\frac{c_{0}}{2}\;.\label{eq:tdde.4}
\end{align}
Substitution of (\ref{eq:td.9}) into (\ref{eq:tdde.4}) yields $c_{0}=\sigma_{0}^{2}$
and 
\begin{align}
k\frac{2t^{\alpha+1}}{\alpha+1} & =\frac{D^{2}t^{2}}{\sigma_{0}^{2}}\;,
\end{align}
which can only be fulfilled by $\alpha=1$, so that 
\begin{align}
k & =\frac{D^{2}}{\sigma_{0}^{2}}\;.
\end{align}
The time-dependent diffusion coefficient $D_{{\rm t}}$ is with (\ref{eq:td.8})
identified as 
\begin{equation}
D_{{\rm t}}:=\frac{D^{2}}{\sigma_{0}^{2}}\, t=u_{0}^{2}\, t=\frac{\hbar^{2}}{4m^{2}\sigma_{0}^{2}}\, t.\label{eq:td.22}
\end{equation}
Finally, Eq.~(\ref{eq:tdde.1}) reads as
\begin{align}
\frac{\partial P}{\partial t} & =\frac{D^{2}t}{\sigma_{0}^{2}}\,\frac{\partial^{2}P}{\partial x^{2}}\label{eq:ballisticDE}
\end{align}
and turns out to be a \textit{ballistic diffusion equation,} defined
by $\alpha=1$, as the special case of an anomalous diffusion where
the diffusion coefficient $D_{{\rm t}}$ grows linearly with time
$t$.

Essentially, the ``decay of the wave packet'' thus simply results
from sub-quantum diffusion with a diffusivity varying in time due
to the particle's changing thermal environment: as the heat initially
concentrated in a narrow spatial domain gets gradually dispersed,
so must the diffusivity of the medium change accordingly.

Now we look at the time $t_{{\rm k}}$ of the kink (Fig.~\ref{fig:td.1}).
The wave packet begins to spread differently at the kink, which is,
according to Eq.~(\ref{eq:td.11}), obviously at that time $t=t_{{\rm k}}$
where the influence of the right term is equal to the left term under
the square root and hence $D^{2}t^{2}\overset{!}{=}\sigma_{0}^{4}$
(i.e., $\sigma=\sqrt{2}\sigma_{0}$). Then we find with (\ref{eq:td.22})
that
\begin{equation}
D_{{\rm t}}=\frac{t}{t_{{\rm k}}}D.\label{eq:td.23}
\end{equation}
As one can see, $t=t_{{\rm k}}$ is the time when $D_{{\rm t}}=D$.
Note that the diffusivity $D$ is constant for all times $t$ and
has to be distinguished from the diffusion coefficient $D_{{\rm t}}$.
In a different approach, one could also start out with the ``exact''
uncertainty relation, $mu_{0}^{2}t_{{\rm k}}=\hbar/2$, with $u_{0}=D/\sigma_{0}$.
This again leads to $D_{{\rm t}}=D^{2}t/\sigma_{0}^{2}=u_{0}^{2}t=t/t_{k}D$.

We recall Boltzmann's relation $\Delta Q=2\omega_{0}\delta S$ \cite{Groessing.2008vacuum,Groessing.2009origin}
between the heat applied to an oscillating system and a change in
the action function $\delta S=\frac{1}{2\pi}\delta\int_{0}^{\tau}E_{{\rm kin}}\d t$,
respectively, providing 
\begin{equation}
\nabla Q=2\omega_{0}\nabla(\delta S)\;.\label{eq:td.24}
\end{equation}
Here, $\delta S$ relates to the momentum fluctuation via 
\begin{equation}
\nabla(\delta S)=\delta\mathbf{p}=:m\mathbf{u}=-\frac{\hbar}{2}\frac{\nabla P}{P}\;,\label{eq:td.25}
\end{equation}
and therefore, with $P=P_{0}\e^{-\delta Q/kT_{0}}$ and $\Delta Q=kT=\hbar\omega$,
\begin{equation}
m\mathbf{u}=\frac{\nabla Q}{2\omega}\;.\label{eq:td.26}
\end{equation}
Using the initial velocity (\ref{eq:td.8}) together with Eq.~(\ref{eq:td.23})
we find
\begin{equation}
D_{{\rm t}}=u_{0}^{2}t=\frac{2}{\hbar}mu_{0}^{2}tD=\frac{2}{\hbar}\frac{(\delta p)^{2}}{2m}tD=\frac{2}{\hbar}\left[\delta S(t)-\delta S(0)\right]D=\frac{\Delta Q(t)}{\hbar\omega}D.\label{eq:td.27}
\end{equation}
Actually, $\delta S(0)=0$, since there are no initial fluctuations.
Substitution of (\ref{eq:td.23}) into (\ref{eq:td.27}) leads then
to
\begin{equation}
D_{{\rm t}}=\frac{t}{t_{{\rm k}}}D=\frac{\Delta Q}{kT}D=-\ln\frac{P(t)}{P(0)}D=-D\left[\ln P(t)-\ln P(0)\right].\label{eq:td.28}
\end{equation}
One can also derive a condition that does not require to know the
diffusion coefficient at $t=0$, 
\begin{equation}
\Delta D=D_{{\rm t}}(t_{2})-D_{{\rm t}}(t_{1})=-D\left[\ln P(t_{2})-\ln P(t_{1})\right],\label{eq:td.29}
\end{equation}
by choosing two arbitrary time steps $t_{1}$ and $t_{2}$ as suggested
in Fig.~\ref{fig:td.1}.

From condition (\ref{eq:td.22}) one can immediately see that
\begin{equation}
\frac{\p D_{{\rm t}}}{\p t}=\frac{D^{2}}{\sigma_{0}^{2}}=\textrm{const.}\label{eq:td.31}
\end{equation}
Thus, one can also rewrite Eq.~(\ref{eq:td.29}) as 
\begin{equation}
\Delta D=D_{{\rm t}}(t_{1})+\frac{D^{2}}{\sigma_{0}^{2}}(t_{2}-t_{1}),\label{eq:td.32}
\end{equation}
which is only valid for equal slit widths $x_{01}=x_{02}$ and thus
$\sigma_{1}=\sigma_{2}$. In order to compute the distribution of
$P(x,t)$, one starts with Eq.~(\ref{eq:td.32}) and takes the local
properties of the diffusivity into account. For given times $t_{1}$
and $t_{2}=t_{1}+\Delta t$ one obtains with (\ref{eq:td.8}), 
\begin{align}
D_{{\rm t}}(x,t_{1}) & =-D\ln P(x,t_{1}),\label{eq:td.36}\\
D_{{\rm t}}(x,t_{2}) & =-D\ln P(x,t_{2})+D_{{\rm t}}(x,t_{1}),\label{eq:td.37}
\end{align}
thereby constituting a rule to numerically compute the distribution
$P(x,t)$.

\section{Finite difference scheme}

Starting with the ballistic diffusion equation (\ref{eq:ballisticDE})
with time-dependent diffusivity $D_{{\rm t}}$ we use an explicit
finite difference forward scheme (cf. \cite{Schwarz.2009numerische}),
\begin{align}
\frac{\partial P}{\partial t} & \rightarrow\frac{1}{\Delta t}\left(P[x,t+1]-P[x,t]\right),\\
\frac{\partial^{2}P}{\partial x^{2}} & \rightarrow\frac{1}{\Delta x^{2}}\left(P[x+1,t]-2P[x,t]+P[x-1,t]\right),
\end{align}
 with 1-dimensional cells. In case $D_{{\rm t}}$ is independent of
$x$, the complete equation after reordering leads to
\begin{equation}
P[x,t+1]=P[x,t]+\frac{D[t+1]\Delta t}{\Delta x^{2}}\left\{ P[x+1,t]-2P[x,t]+P[x-1,t]\right\} \label{eq:tdde.1.2.3}
\end{equation}
with space $x$ and time $t$, and initial Gaussian distribution $P(x,0)$
with standard deviation $\sigma_{0}$.

As can be seen, calculation of a cell's value at time $t$ only depends
on cell values at the previous time. The time-dependence of the diffusion
coefficient can also be calculated without any knowledge of neighboring
cells. The diffusion coefficient represents the underlying physics
of the current cell and is calculated for the evaluated time step
$t+1$.

The stability condition for the scheme of Eq.~(\ref{eq:tdde.1.2.3})
is that 
\begin{equation}
\left|\frac{D_{{\rm t}}\Delta t}{\Delta x^{2}}\right|\leq\frac{1}{2}\label{eq:tdde.1.4.1}
\end{equation}
be satisfied for all values of the cells $[x,t]$ in the domain of
computation. The general procedure is that one considers each of the
\textit{frozen coefficient problems} arising from the scheme. The
frozen coefficient problems are the constant coefficient problems
obtained by fixing the coefficients at their values attained at each
point in the domain of the computation (cf.~\cite{Strikwerda.2004finite}).
Substituting Eq.~(\ref{eq:td.22}) into (\ref{eq:tdde.1.4.1}) leads
to 
\begin{equation}
\Delta t\leq\frac{\Delta x^{2}\sigma_{0}^{2}}{2D_{{\rm t}}^{2}t}\:.\label{eq:tdde.1.4.2}
\end{equation}
This shows that the finite difference scheme (\ref{eq:tdde.1.2.3})
is suitable to solve the ballistic diffusion equation~(\ref{eq:ballisticDE})
as long as the spreading is not too big. Beyond that, the computations
are no longer economically practical due to the necessarily enormous
number of cells. Then, one has to replace the explicit scheme~(\ref{eq:tdde.1.2.3})
by an implicit scheme, for example, which has less stringent restrictions
on the stability conditions, but needs linear equation solvers instead.
For our simulations we employed the explicit scheme introduced above
as well as implicit schemes with an open source software for numerical
computation, Scilab \cite{Scilab}, on a standard personal computer.

\section{The connection to wave mechanics: The double-slit experiments with
different slit widths}

For a more generalized picture, we now take a closer look at the double-slit
experiment.
\begin{figure}[th]
\centering{}\includegraphics{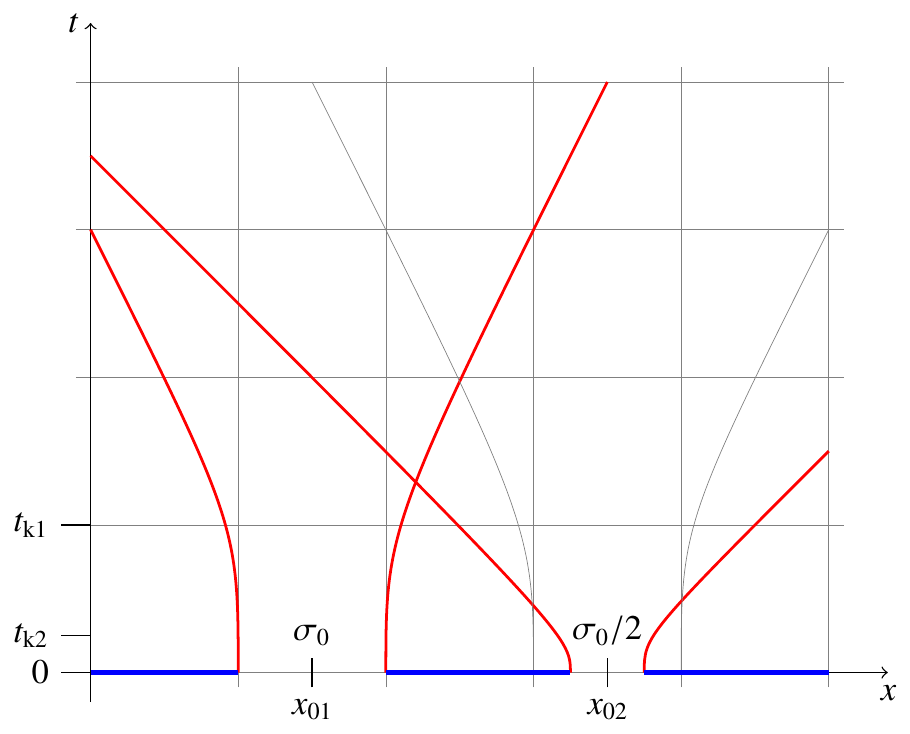}\caption{Sketch of a double-slit with two different widths and Bohm-type trajectories
(and same-widths scenario indicated by gray lines)\label{fig:td.2}}
\end{figure}
Consider a scenario as shown in Fig.~\ref{fig:td.2} with two slits
of different widths. We assume the initial Gaussians passing through
a slit have a standard deviation value matching the slit width, e.g.,
$\sigma_{01}=\sigma_{0}$ and $\sigma_{02}=\sigma_{0}/2$, respectively,
with $\sigma_{0i}$ then being also the width of slit~$i$. The resulting
Bohm-type trajectories of the two decaying Gaussians are sketched
in Fig.~\ref{fig:td.2} with red lines. Thus 
\begin{equation}
t_{{\rm k}2}=t_{{\rm k}1}/4
\end{equation}
while the spreading is doubled (compare with the grayed out spreading
of slit~2 for the case of $\sigma_{0}$ for both slits). According
to Eq.~(\ref{eq:td.22}), the diffusion coefficients of the two slits
yield 
\begin{equation}
D_{{\rm t},1}(t)=\frac{D^{2}t}{\sigma_{02}^{2}}\neq D_{{\rm t},2}(t)=\frac{D^{2}t}{\sigma_{01}^{2}},\quad t>0.
\end{equation}

The advantage of Eq.~(\ref{eq:td.28}) lies in it's local fit due
to its dependence on $P(x,t)$ instead on $\sigma(t)$, since the
latter is just a global statistical value of too less local relevance.
For the general case, we have to deal with a diffusion coefficient
$D_{{\rm t}}(x,t)$ further on. The time-dependent diffusion equation
reads then 
\begin{equation}
\frac{\partial P}{\partial t}=\frac{\p}{\p x}\left(D_{{\rm t}}(x,t)\frac{\partial P}{\partial x}\right)\;.\label{eq:td.35}
\end{equation}

We have now all the tools necessary to consider the inclusion of wave
mechanics in our model. We define the phase as
\begin{equation}
\varphi=S/\hbar\label{eq:td.38}
\end{equation}
with the general action $S$. Identifying $v_{{\rm tot}}$ of Eq.~(\ref{eq:td.20})
with 
\begin{equation}
v_{{\rm tot}}=\frac{\nabla S}{m}\label{eq:td39}
\end{equation}
we find for the action
\begin{equation}
S=\int mv_{{\rm tot}}(t)\d x-\int E\d t=m\int\left(v+\frac{u_{0}^{2}t}{\sigma_{0}^{2}+u_{0}^{2}t^{2}}\,\xi(t)\right)\d x-\int E\d t,\label{eq:td.40}
\end{equation}
with $E$ being the system's total energy. As $v$ does not depend
on $x$ we can solve the first integral, and for the conservative
case also the second integral, providing with (\ref{eq:td.12})
\begin{equation}
S=mvx+\frac{mu_{0}^{2}}{2}\left(\frac{\xi(t)}{\sigma(t)}\right)^{2}t-Et=mvx+\frac{mu_{0}^{2}}{2}\left(\frac{\xi(0)}{\sigma_{0}}\right)^{2}t-Et.\label{eq:td.41}
\end{equation}
Here, the action $S$ \textit{along a trajectory} is given by the
sum of the usual momentum-related term and a term depending on the
kinetic energy, or kinetic temperature, respectively, of the ``heated
up'' environment, weighted by a factor that solely depends on a particular
trajectory indicated by the initial location $\xi(0)$ in the Gaussian.

Finally, we rewrite the phase with the help of Eqs.~(\ref{eq:td.38})
and (\ref{eq:td.17}) as
\begin{equation}
\varphi=\frac{1}{\hbar}\left[mvx+\frac{mu_{0}^{2}}{2}\left(\frac{\xi(0)}{\sigma_{0}}\right)^{2}t-Et\right]=\frac{1}{\hbar}\left[mvx+\frac{mu_{0}^{2}}{2}\left(\frac{x-vt}{\sigma(t)}\right)^{2}t-Et\right].\label{eq:td.42}
\end{equation}
The expression containing $\xi(0)$ indicates a phase $\varphi$ along
a trajectory, while the r.h.s.~sticks to our coordinate system and
is thus the better choice to do interference calculations.

Instead of following just one Gaussian, we extend our simulation scheme
to include two possible paths of a particle which eventually cross
each other. For this, we use two Gaussians approaching each other.
Following our earlier approach in \cite{Groessing.2012doubleslit}
we simulate a double-slit experiment by independent numerical computation
of two Gaussian wave packets with total distribution given by
\begin{equation}
P_{{\rm tot}}:=P_{1}+P_{2}+2\sqrt{P_{1}P_{2}}\cos\varphi_{12}.\label{eq:td.43}
\end{equation}
Since each Gaussian has its own phase (\ref{eq:td.42}) we are free
to add a phase shifter $\Delta\varphi$ for one of the slits of the
two-slit experiment, say slit 1, which modifies the phase to
\begin{equation}
\varphi{}_{1}=\frac{S_{1}}{\hbar}+\Delta\varphi\label{eq:td.46}
\end{equation}
and yields for the phase difference

\begin{align}
\varphi_{12}=\varphi_{2}-\varphi{}_{1}= & \frac{m}{\hbar}\left[\vphantom{\intop}v_{2}(x-x_{02})-v_{1}(x-x_{01})\right]\label{eq:td.44}\\
 & +\frac{mt}{2\hbar}\left[\frac{u_{02}^{2}(x-x_{02}-v_{2}t)^{2}}{\sigma_{2}^{2}(t)}-\frac{u_{01}^{2}(x-x_{01}-v_{1}t)^{2}}{\sigma_{1}^{2}(t)}\right]-\Delta\varphi.\nonumber 
\end{align}
The two slits at positions $x_{01}$ and $x_{02}$ have different
slit widths and hence different parameters, $\sigma_{01}$, $\sigma_{1}$,
$u_{01}$ and $\sigma_{02}$, $\sigma_{2}$, $u_{02}$, respectively,
as illustrated by the red trajectories in Fig.~\ref{fig:td.2} for
the example of $\sigma_{01}=\sigma_{0}$ and $\sigma_{02}=\sigma_{0}/2$,
respectively. One can observe several characteristics of the averaged
particle trajectories, which, just because of the averaging, are identical
with the Bohmian trajectories. As one can see, the phase difference
(\ref{eq:td.44}) is at any time defined for the whole domain, and
hence $\varphi_{12}$ is intrinsically nonlocal. 

Finally, we recall our derivation of the total average density current
\cite{Groessing.2012doubleslit,Grossing.2012quantum}, i.e., the most
general expression (including weights $P_{i}$) for our ``path excitation
field'',
\begin{equation}
J_{{\rm tot}}=P_{1}v_{1}+P_{2}v_{2}+\sqrt{P_{1}P_{2}}\left(v_{1}+v_{2}\right)\cos\varphi_{12}+\sqrt{P_{1}P_{2}}\left(u_{1}-u_{2}\right)\sin\varphi_{12},\label{eq:td.45}
\end{equation}
where
\begin{equation}
v_{{\rm tot}}=\frac{J_{{\rm tot}}}{P_{{\rm tot}}}\:,
\end{equation}
with osmotic velocities $u_{i}$ of Eq.~(\ref{eq:td.7}) and total
velocities $v_{i}$ of Eq.~(\ref{eq:td39}) applied to both slits,
1 and 2, and with the phases~(\ref{eq:td.44}). Note that the last
term on the r.h.s~in Eq.~(\ref{eq:td.45}) is termed ``entangling
current'' $J_{\rm e}$ by us \cite{Groessing.2012doubleslit}, which is of a genuinely
``quantum'' nature in that the velocities $u_{i}$ are generally
entangled with the velocities $v_{i}$.

\section{Simulation results}

\begin{figure}[th]
\subfloat[equal slit widths\label{fig:td.3a}]{\centering{}\includegraphics{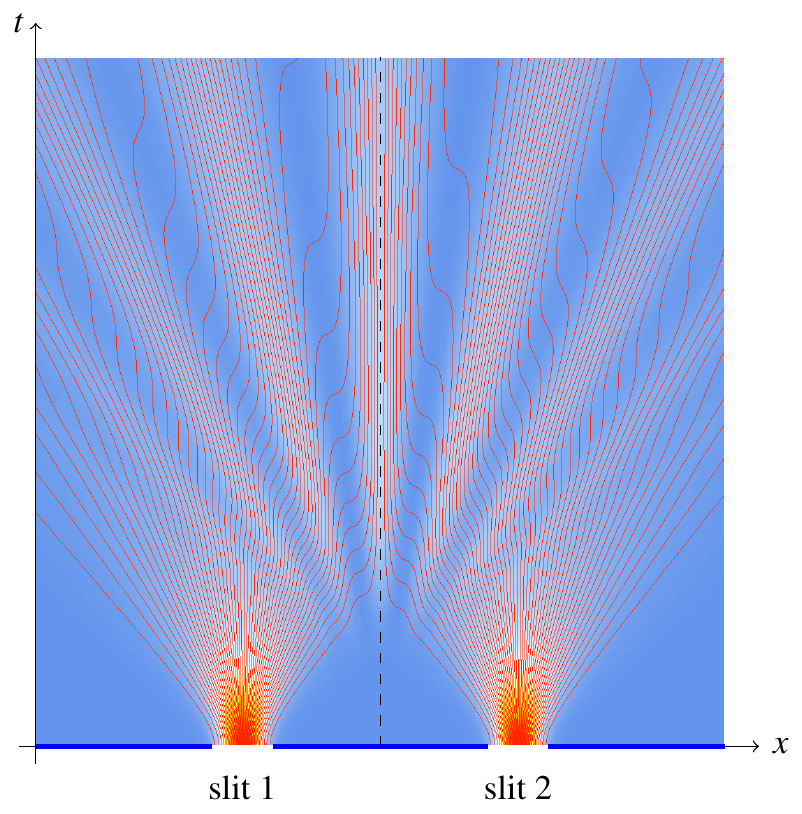}}\subfloat[$\sigma_{01}=2\sigma_{02}$\label{fig:td.3b}]{\centering{}\includegraphics{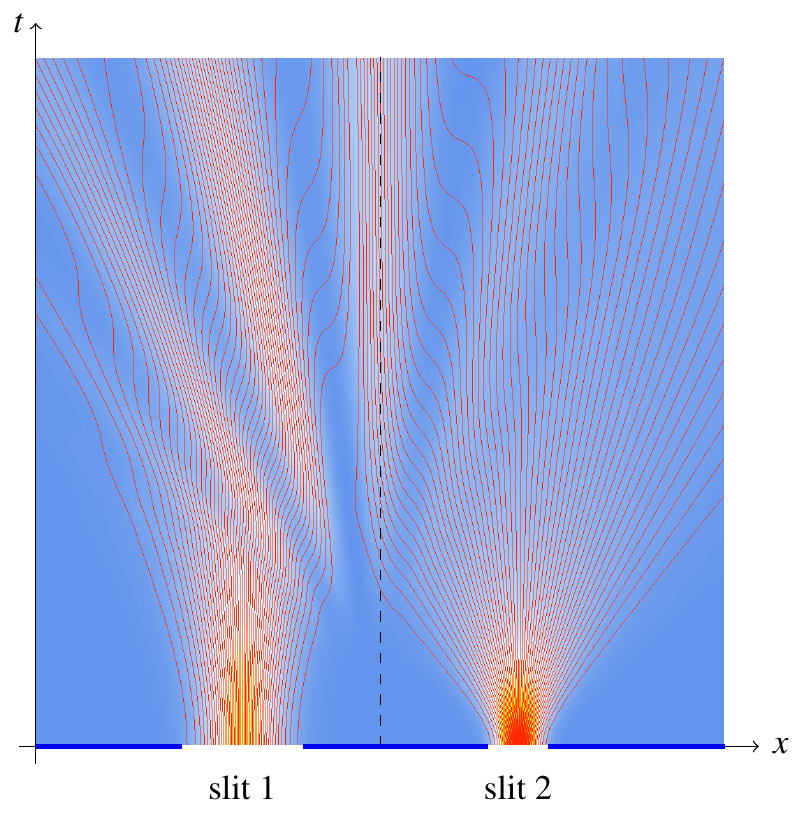}}\caption{Classical computer simulation of the interference pattern with different
slit widths: intensity distribution with increasing intensity from
white through yellow and orange, with trajectories (red) for two Gaussian
slits ($v_{x,1}=v_{x,2}=0$)\label{fig:td.3}}
\end{figure}

\begin{figure}[!tbh]
\subfloat[Probability density $P$\label{fig:td.4a}]{\centering{}\includegraphics{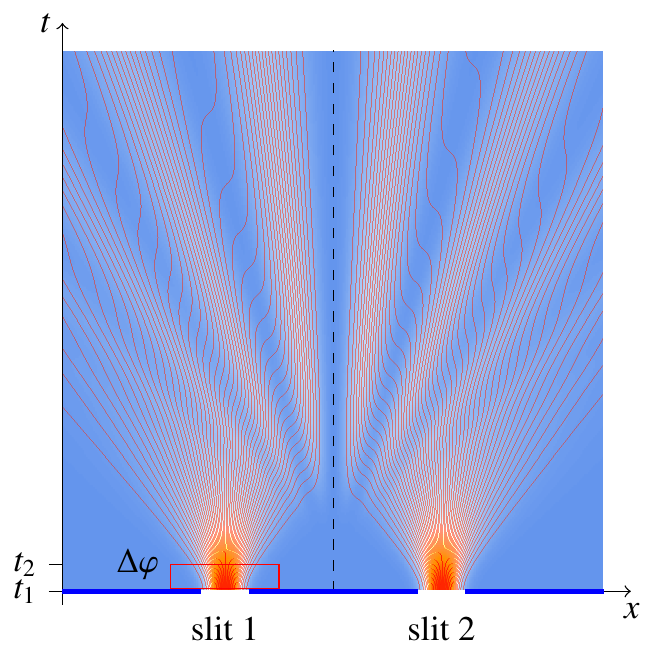}}\subfloat[Phase shift\label{fig:td.4b}]{\centering{}\includegraphics{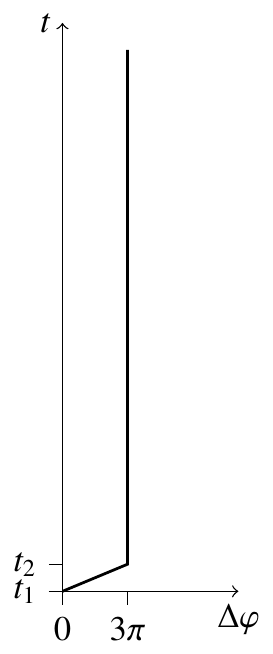}}\subfloat[Entangling current~$J_{{\rm e}}$\label{fig:td.4c}]{\centering{}\includegraphics{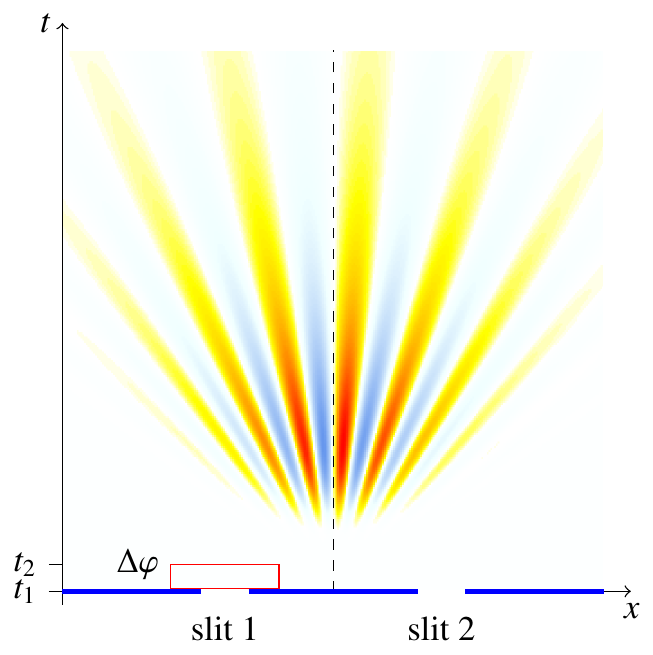}}\caption{Classical computer simulation as in Fig.~\ref{fig:td.3a}, but with
additional phase shift $\Delta\varphi=3\pi$ accumulated during the
time interval between $t_{1}$ and $t_{2}$ at slit~1\label{fig:td.4}}
\end{figure}

\begin{figure}[!tbh]
\subfloat[Probability density $P$\label{fig:td.5a}]{\centering{}\includegraphics{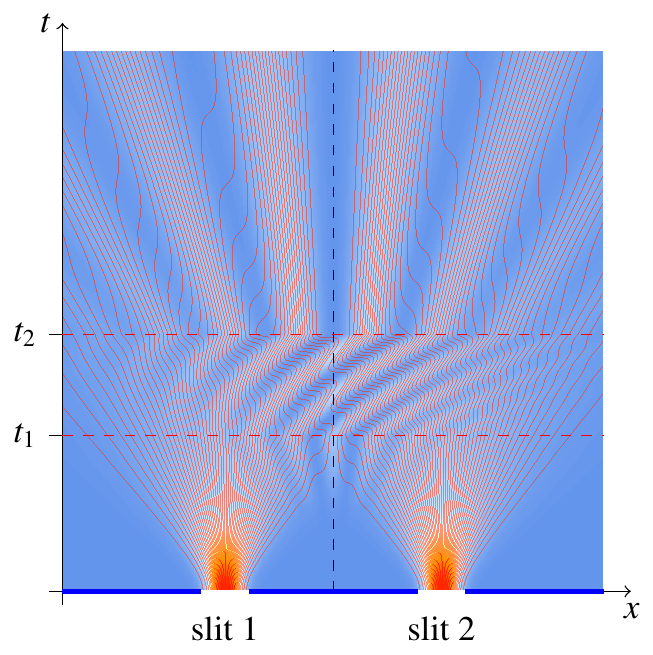}}\subfloat[Phase shift\label{fig:td.5b}]{\centering{}\includegraphics{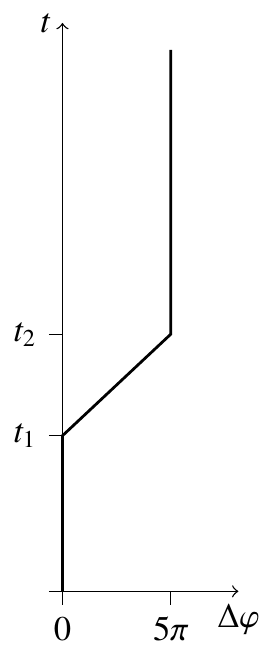}}\subfloat[Entangling current~$J_{{\rm e}}$\label{fig:td.5c}]{\centering{}\includegraphics{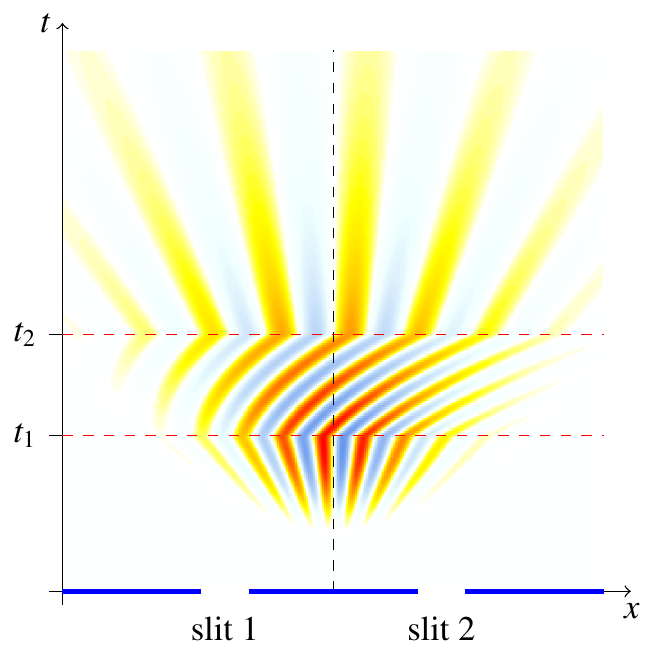}}\caption{Classical computer simulation as in Fig.~\ref{fig:td.4}, but with
different times $t_{i}$ and with accumulated additional phase $\Delta\varphi=5\pi$.
This results in the same distributions of $P$ and $J_{{\rm e}}$
for times $t>t_{2}$ and shows the effect of the shifting of the interference
fringes more clearly than Fig.~\ref{fig:td.4}\label{fig:td.5}}
\end{figure}

In Figs.~\ref{fig:td.3} to \ref{fig:td.5}, the graphical results
of a classical computer simulation of the interference pattern in
double-slit experiments are shown, including the trajectories. In
Fig.~\ref{fig:td.3a} the maximum of the intensity is distributed
along the symmetry line exactly in the middle between the two slits.
\cite{Groessing.2012doubleslit}

In the examplary figures, trajectories according to Eq.~(\ref{eq:td.43})
for the two Gaussian slits are shown. The interference hyperbolas
for the maxima characterize the regions where the phase difference
$\varphi=2n\pi$, and those with the minima lie at $\varphi=(2n+1)\pi$,
$n=0,\pm1,\pm2,\ldots$ Note in particular the ``kinks'' of trajectories
moving from the center-oriented side of one relative maximum to cross
over to join more central (relative) maxima. In our classical explanation
of double slit interference, a detailed ``micro-causal'' account
of the corresponding kinematics can be given. The trajectories are
in full accordance with those obtained from the Bohmian approach,
as can be seen by comparison with \cite{Holland.1993,Bohm.1993undivided,Sanz.2009context,Sanz.2012trajectory},
for example.

We use the same double-slit arrangements in Figs.~\ref{fig:td.4}
and \ref{fig:td.5}, but include a phase shifter affecting the current
from slit~1, as sketched by the dashed red lines or rectangles on
the left hand side, respectively. Even though the total applied phase
shift is either $3\pi$ or $5\pi$ in Figs.~\ref{fig:td.4a} and
\ref{fig:td.5a}, respectively, one recognizes the effective phase
difference of $\Delta\varphi_{{\rm mod}}=\Delta\varphi\mod2\pi=\pi$
in each case, which eventually results in equal shifts of the interference
fringes. By comparing with Fig.~\ref{fig:td.3a} we now observe a
minimum of the resulting distribution along the central symmetry line,
in full accordance with the Aharonov-Bohm effect, independently of
the times $t_{1}$ and $t_{2}$ during which the shift has been applied.

To bring out the shifting of the interference fringes more clearly,
we apply in Fig.~\ref{fig:td.5} the phase shift in between the indicated
times $t_{1}$ and $t_{2}$, respectively. Note that the phase shift
applied only to a single slit's current at a time $t$ when the decaying
Gaussians are already overlapping (Fig.~\ref{fig:td.5}) is shown
here for didactic reasons only. In this highly idealized scenario,
then, one can see an illustration of the immediate effectiveness of
$\Delta\varphi$ over the whole domain according to Eq.~(\ref{eq:td.44}),
i.e., of the nonlocality of the relative phase.


To conclude, we have in this paper provided a detailed description
of the velocity fields involved in the analytical calculations as
well as the computer simulations illustrating Gaussian dispersion
and interference at a double slit. We have arrived at an expression
for the local value of the phase, Eq.~(\ref{eq:td.42}), which made
it possible also to extend our previous model to slit systems with
independently variable slit widths. With the computer simulations
of the latter, the nonlocal nature of the relative phase can be clearly
demonstrated.

\section*{Acknowledgements}

We thank Hans-Thomas Elze for pointing out to us that Gaussian
dispersion can be obtained via classical modeling with Liouville path
integrals, Jan Walleczek for enlightening discussions
on numerous related issues, and the Fetzer Franklin Found for partial support.

\providecommand{\href}[2]{#2}
\begingroup\raggedright
\endgroup


\begin{thebibliography}{10}

\bibitem{Groessing.2010emergence}
G.~Gr\"{o}ssing, S.~Fussy, J.~Mesa~Pascasio, and H.~Schwabl, ``Emergence and
  collapse of quantum mechanical superposition: Orthogonality of reversible
  dynamics and irreversible diffusion,''
  \href{http://dx.doi.org/10.1016/j.physa.2010.07.017}{{\em Physica A}
  {\bfseries 389} (2010) 4473--4484},
  \href{http://arxiv.org/abs/1004.4596}{{\ttfamily {arXiv}:1004.4596
  [quant-ph]}}.

\bibitem{Groessing.2012doubleslit}
G.~Gr\"{o}ssing, S.~Fussy, J.~Mesa~Pascasio, and H.~Schwabl, ``An explanation
  of interference effects in the double slit experiment: Classical trajectories
  plus ballistic diffusion caused by zero-point fluctuations,''
  \href{http://dx.doi.org/10.1016/j.aop.2011.11.010}{{\em Ann. Phys.}
  {\bfseries 327} (2012) 421--437},
  \href{http://arxiv.org/abs/1106.5994}{{\ttfamily {arXiv}:1106.5994
  [quant-ph]}}.

\bibitem{Couder.2012probabilities}
Y.~Couder and E.~Fort, ``Probabilities and trajectories in a classical
  wave-particle duality,''
  \href{http://dx.doi.org/10.1088/1742-6596/361/1/012001}{{\em J. Phys.: Conf.
  Ser.} {\bfseries 361} (2012) 012001}.

\bibitem{Holland.1993}
P.~R. Holland, {\em The Quantum Theory of Motion: An account of the de
  Broglie-Bohm causal interpretation of quantum mechanics}.
\newblock Cambridge University Press, Cambridge, 1993.

\bibitem{Elze.2011general}
H.~T. Elze, G.~Gambarotta, and F.~Vallone, ``General linear dynamics - quantum,
  classical or hybrid,''
  \href{http://dx.doi.org/10.1088/1742-6596/306/1/012010}{{\em J. Phys.: Conf.
  Ser.} {\bfseries 306} (2011) 012010}.

\bibitem{DeBroglie.1960book}
L.~V. P.~R. de~Broglie, {\em Non-Linear Wave Mechanics: A Causal
  Interpretation.}
\newblock Elsevier, Amsterdam, 1960.

\bibitem{Hall.2002schrodinger}
M.~J.~W. Hall and M.~Reginatto, ``Schr\"{o}dinger equation from an exact
  uncertainty principle,''
  \href{http://dx.doi.org/10.1088/0305-4470/35/14/310}{{\em J. Phys. A: Math.
  Gen.} {\bfseries 35} (2002) 3289--3303}.

\bibitem{Mesa.2012classical}
J.~Mesa~Pascasio, S.~Fussy, H.~Schwabl, and G.~Gr\"{o}ssing, ``Classical
  simulation of double slit interference via ballistic diffusion,''
  \href{http://dx.doi.org/10.1088/1742-6596/361/1/012041}{{\em J. Phys.: Conf.
  Ser.} {\bfseries 361} (2012) 012041},
  \href{http://arxiv.org/abs/1205.4521}{{\ttfamily {arXiv}:1205.4521
  [quant-ph]}}.

\bibitem{Groessing.2008vacuum}
G.~Gr\"{o}ssing, ``The vacuum fluctuation theorem: Exact schr\"{o}dinger
  equation via nonequilibrium thermodynamics,''
  \href{http://dx.doi.org/10.1016/j.physleta.2008.05.007}{{\em Phys. Lett. A}
  {\bfseries 372} (2008) 4556--4563},
  \href{http://arxiv.org/abs/0711.4954v2}{{\ttfamily {arXiv}:0711.4954v2
  [quant-ph]}}.

\bibitem{Groessing.2009origin}
G.~Gr\"{o}ssing, ``On the thermodynamic origin of the quantum potential,''
  \href{http://dx.doi.org/10.1016/j.physa.2008.11.033}{{\em Physica A}
  {\bfseries 388} (2009) 811--823},
  \href{http://arxiv.org/abs/0808.3539v1}{{\ttfamily {arXiv}:0808.3539v1
  [quant-ph]}}.

\bibitem{Schwarz.2009numerische}
H.~R. Schwarz and N.~K\"{o}ckler, {\em Numerische Mathematik}.
\newblock {Vieweg+Teubner}, 7~ed., 2009.

\bibitem{Strikwerda.2004finite}
J.~C. Strikwerda, {\em Finite Difference Schemes and Partial Differential
  Equations}.
\newblock {SIAM}, 2~ed., 2004.

\bibitem{Scilab}
Scilab, ``Free open source software for numerical computation.''
  Http://www.scilab.org/, 2012.

\bibitem{Grossing.2012quantum}
G.~Gr\"{o}ssing, S.~Fussy, J.~Mesa~Pascasio, and H.~Schwabl, ``The quantum as
  an emergent system,''
  \href{http://dx.doi.org/10.1088/1742-6596/361/1/012008}{{\em J. Phys.: Conf.
  Ser.} {\bfseries 361} (2012) 012008},
  \href{http://arxiv.org/abs/1205.3393}{{\ttfamily {arXiv}:1205.3393
  [quant-ph]}}.

\bibitem{Bohm.1993undivided}
D.~Bohm and B.~J. Hiley, {\em The undivided universe: An ontological
  interpretation of quantum theory}.
\newblock Routledge, London, 1993.

\bibitem{Sanz.2009context}
A.~S. Sanz and F.~Borondo, ``Contextuality, decoherence and quantum
  trajectories,'' \href{http://dx.doi.org/10.1016/j.cplett.2009.07.061}{{\em
  Chem. Phys. Lett.} {\bfseries 478} (2009) 301--306},
  \href{http://arxiv.org/abs/0803.2581}{{\ttfamily {arXiv}:0803.2581
  [quant-ph]}}.

\bibitem{Sanz.2012trajectory}
A.~S. Sanz and S.~Miret-Art\'{e}s, {\em A Trajectory Description of Quantum
  Processes. I. Fundamentals. A Bohmian Perspective}, vol.~850 of {\em Lecture
  Notes in Physics}.
\newblock Springer, 2012.

\end{thebibliography}
\end{document}